\documentclass[journal,transmag]{IEEEtran}
\usepackage{mathtools}
\usepackage{float}
\usepackage{caption}
\usepackage{graphicx} 
\usepackage[colorinlistoftodos]{todonotes}
\graphicspath{ {images/} }

\usepackage[linesnumbered,ruled,vlined]{algorithm2e} 

\usepackage[utf8]{inputenc}
\usepackage[english]{babel}
\usepackage{amsthm,amssymb,amsmath}
\usepackage{nicematrix}
\usepackage{optidef}

\theoremstyle{definition}
\newtheorem{lemma}{Lemma}
\newtheorem{corollary}{Corollary}

\newtheorem{example}{Example}

\IEEEoverridecommandlockouts                              

\title{ 
On Convolutional Precoding in PAC Codes
}

\author{
\IEEEauthorblockN{Mohammad Rowshan, {\em Student Member, IEEE} and Emanuele Viterbo, {\em Fellow, IEEE}}
 \thanks{M. Rowshan and E. Viterbo are with the  Department of Electrical and Computer Systems Engineering (ECSE), Monash University, Melbourne, VIC3800, Australia. E-mail: \{mohammad.rowshan, emanuele.viterbo\}@monash.edu. These authors' work was supported by the Australian Research Council under Discovery Project ARC DP160100528.}
}

\begin{document}

\maketitle
\thispagestyle{empty}
\pagestyle{empty}

\begin{abstract}
Polarization-adjusted convolutional (PAC) codes are special concatenated codes in which we employ a one-to-one convolutional transform as a precoding step before the polar transform. In this scheme, the polar transform (as a mapper) and the successive cancellation process (as a demapper) present a synthetic vector channel to the convolutional transformation. The numerical results in the literature show that this concatenation improves the weight distribution of polar codes which justifies the superior error correction performance of PAC codes relative to polar codes. In this work, we explicitly show why the convolutional precoding reduces the number of minimum-weight codewords. Further analysis exhibits where the precoding stage is not effective. 
Then, we recognize weaknesses of the convolutional precoding which are unequal error protection (UEP) of the information bits due to rate profiling and lack of cross-segmental convolution. Finally, we assess the possibility of improving the precoding stage by proposing some irregular convolutional precodings.
\end{abstract}


\section{Introduction}
\label{sec:intro}
Polar codes proposed by Ar\i kan in \cite{arikan} are the first class of  channel codes with an  explicit construction that was proven to achieve the symmetric (Shannon) capacity of a binary-input discrete memoryless channel (BI-DMC) using a low-complexity successive cancellation (SC) decoder (SCD). Nevertheless, 
the error correction performance of finite-length polar codes under SCD is not satisfactory due to the existence of partially polarized channels. 

\begin{figure}
    \centering
    \includegraphics[width=1\columnwidth]{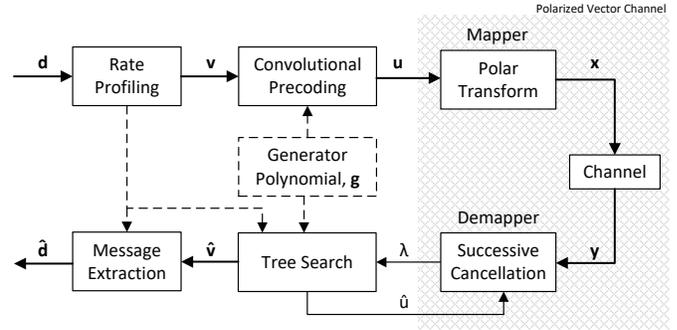}
    \caption{\vspace{-5mm} PAC Coding Scheme} 
    \label{fig:PAC_scheme}
\end{figure}

Recently in \cite{arikan2}, Ar\i kan proposed a concatenation of a convolutional precoding/transform with the polarization transform \cite{arikan} where a message is first encoded using a convolutional transform and then transmitted over polarized synthetic channels as shown in Fig.~\ref{fig:PAC_scheme}. These codes are called ``polarization-adjusted convolutional (PAC) codes". 
It was shown in \cite{li2} that a properly designed pre-transformation, such as a convolutional transform, can improve the distance properties of polar codes.  Hence, as it was shown in \cite{rowshan-pac1}, PAC codes can outperform polar codes without CRC concatenation for short code lengths and CRC aided PAC codes can outperform CRC-aided polar codes for long code lengths. In \cite{rowshan-pac1}, we also studied the implementation of tree search algorithms including the conventional list decoding, stack decoding, and complexity-efficient Fano decoding for PAC codes. The list Viterbi decoding was adapted to PAC codes in \cite{rowshan-lva1}. 

In this work, we analyse the impact of precoding stage on the distance properties of polar codes. We also show where the precoding stage is not effective depending on the rate profile. Then, we recognize the weaknesses of the convolutional precoding in PAC codes and propose some schemes to mitigate them. Finally, we assess the impact of the proposed schemes on the distance properties and FER performance.


\section{Preliminaries}\label{sec:prelim}
Polarization-adjusted convolutional (PAC) codes are denoted by {\em PAC}$(N,K,\mathcal{A},\mathbf{g})$, where $N = 2^n$ is the length of the PAC code. A rate profiler first maps the $K$ information bits to $N$ bits. Then, the convolutional transform (with polynomial coefficients vector $\mathbf{g}$) scrambles the resulting $N$ bits before feeding them to the classical polar transform (Fig.~\ref{fig:PAC_scheme}). 
The information bits $\mathbf{d}=[d_0,d_1,...,d_{K-1}]$ are interspersed with $N-K$ zeros and mapped to the vector $\mathbf{v}=[v_0,v_1,...,v_{N-1}]$ using a rate-profile which defines the code construction. The rate-profile is defined by the index set $\mathcal{A}\subseteq \{0,\ldots, N-1\}$, where the information bits appear in $\mathbf{v}$. This set can be defined as the indices of sub-channels in the polarized vector channel with high reliability. 
These sub-channels are called {\em good channels}. The bit values in the remaining positions $\mathcal{A}^c$ in $\mathbf{v}$ are set to 0.  

\begin{figure}
    \centering
    \vspace{-3mm}
    \includegraphics[width=0.7\columnwidth]{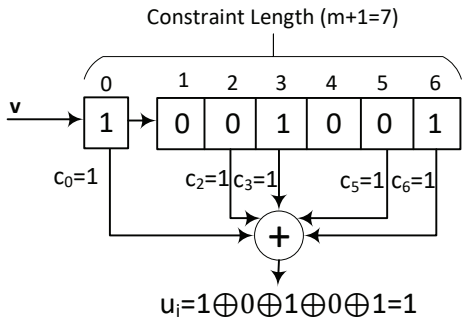}
    \caption{An example of convolution operation using a shift-register}
    \label{fig:conv}
    \vspace{-5pt}
\end{figure}

The input vector $\mathbf{v}$ is transformed to vector $\mathbf{u}=[u_0,\ldots,u_{N-1}]$ as $u_i = \sum_{j=0}^m g_j v_{i-j}$ using the binary generator polynomial of degree $m$, with coefficients $\mathbf{g}=[g_0,\ldots,g_m]$. 
This convolutional transformation combines $m$ previous input bits stored in a shift register with the current input bit $v_i$ as shown in Fig. \ref{fig:conv} to calculate $u_i$. The parameter $m+1$, in bits, is called the {\em constraint length} of the convolutional code. 

Equivalently, the convolution operation can be represented in the form of Toeplitz matrix where the rows of a {\em generator matrix} $G$ are formed by shifting the vector $\mathbf{g} = (g_0,g_1,\ldots g_m)$ one element at a row as shown in (\ref{eq:conv_gen}). 
\begin{equation}
\label{eq:conv_gen}
\mathbf{G}=\begin{bNiceMatrix}
g_0 & g_1 & \Cdots & g_m & 0 & \Cdots & & 0 \\
0 & g_0 & g_1 & \Cdots & g_m & & & \\
\Vdots & \Ddots & \Ddots & \Ddots & & \Ddots & \Ddots &\Vdots \\
 & & & & & & & 0 \\
 & & & & g_0 & g_1 & \Cdots & g_m \\
\Vdots & & & & \Ddots & \Ddots & \Ddots &\Vdots \\
 & & & & & & g_0 & g_1 \\
0 & \Cdots & & & & & 0 & g_0
\end{bNiceMatrix}
\end{equation}
Note that $g_0$ by convention is always $g_0=1$, hence it is an upper-triangular matrix. 
Then, we can obtain $u$ by matrix multiplication as $\mathbf{u}=\mathbf{v}\mathbf{G}$. 
As a result of this pre-transformation, $u_i$ for $i\in\mathcal{A}^c$ are no longer frozen as in polar codes.

Since this convolutional transformation is one-to-one, 
it is not equivalent to a classical generator matrix of convolutional codes. The rate-profiling process performed before the convolutional transformation creates the redundancy by inserting $N-K$ zeros in the length-$K$ input sequence $\mathbf{d}$.  

Finally, as shown in Fig.~\ref{fig:PAC_scheme}, vector $\mathbf{u}$ is mapped to vector $\mathbf{x}$ ($\mathbf{x}=\mathbf{u}\mathbf{P}_n$) by the polar transform $\mathbf{P}_n=\mathbf{P}^{\otimes n}$ defined as the $n$-th Kronecker power of  
$\mathbf{P} = 
{\footnotesize \begin{bmatrix}
1 & 0 \\
1 & 1
\end{bmatrix} }$.

\section{Minimum-weight Codewords in PAC Codes}\label{sec:weight_reduction}
Enumeration of minimum Hamming weight codewords of PAC codes in \cite{rowshan-pac1} showed that they have a significantly less number of min-weight codewords in comparison with polar codes. Let $A_{d_{min}}$ denote the number of codewords with minimum Hamming weight codewords, or in short min-weight codewords. Table~\ref{tbl:1} compares min-weight codewords of polar codes and PAC codes. In this work, the method discussed in \cite{rowshan-pac1} was employed with $L=2^{19}$ to obtain $A_{d_{min}}$.

\begin{table}[h!]
\centering
\caption{The (approximate) number of min-weight codewords, $A_{d_{min}}$, with RM-polar rate profile}
\begin{tabular}{|c |c c c|} 
 \hline
   & (128,32,16) & (128,64,16) & (128,96,8) \\ 
 \hline
 Polar Codes & 56 & 94488 & 74288 \\ 
 PAC Codes & 56 & 3120 & 13904 \\
 \hline
   & (64,16,16) & (64,32,8) & (64,48,4) \\ 
 \hline
 Polar Codes & 364 & 664 & 432 \\ 
 PAC Codes & 236 & 472 & 320 \\
 \hline
\end{tabular}
\label{tbl:1}
\end{table}

As can be seen, $A_{d_{min}}=A_{16}= 94488$ for the polar code (128,64,16) constructed with RM rate profile, whereas $A_{16}\approx 3120$ is much smaller for the PAC code (128,64,16) with the same rate profile. In this section, we discuss the reason behind this significant reduction. 

It was shown in \cite{li2} by example 
that a properly designed upper-triangular matrix $\mathbf{G}$ in general may remove some of the bit-patterns with minimum Hamming weight from the codebook as a result of $\mathbf{G}\mathbf{P}_n$ matrix multiplication in $\mathbf{v}\big(\mathbf{G}\mathbf{P}_n\big)$. 

In this work, we show how convolutional precoding, i.e., $\mathbf{v}\mathbf{G}$ matrix multiplication in $\big(\mathbf{v}\mathbf{G}\big)\mathbf{P}_n$, can avoid generating some of the minimum weight codewords available in the codebook of polar codes generated by $\mathbf{v}\mathbf{P}_n$.

First, let us look at the process that min-weight codewords are generated. The rows of $\mathbf{P}_n$ in $\mathcal{A}$ with min-weight are individually considered as min-weight codewords. The other min-weight codewords are generated by the combination of two or more rows. Here, we just show it for the case of individual row codewords, as the other cases follow the same concept.

We define the cosets resulting from combining a min-weight row at coordinate $i$ with possibly other rows with indices larger than $i$ as 

\begin{equation} \label{eq:coset1}
C(0_0^{i-1},1) = \mathbf{g}_i \oplus \bigoplus_{k\in I} \mathbf{g}_k 
\end{equation} 
where $I\subset \{j|j>i\}$. The following lemma defines a lower bound for the weight of codewords in the coset $C(0_0^{i-1},1)$. The notation $w(.)$ is used for the Hamming weight of vectors.

\begin{lemma} \label{lma:1}
The weight of any codeword in the coset $C(0_0^{i-1},1)$ is 
$w\big(C(0_0^{i-1},1)\big)\geq w(\mathbf{g}_i)$.
\end{lemma}
\begin{proof}
This can be shown by mathematical induction (see  \cite[Corollary 1]{li2}).
\end{proof}

\begin{figure}
    \centering
    \vspace{-3mm}
    \includegraphics[width=0.8\columnwidth]{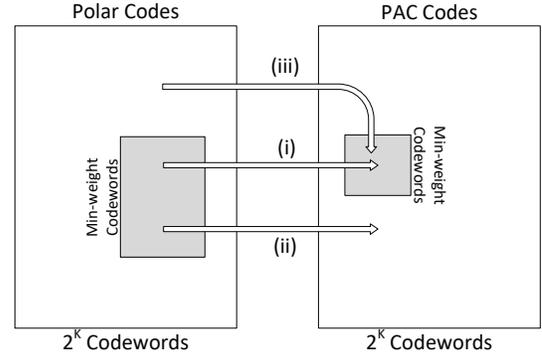}
    \caption{Mapping of min-weight codewords in the codebook of polar codes to PAC codes'.}
    \label{fig:mapping}
\end{figure}

Now, given a polar code with  length $N$, the index set $\mathcal{A}$ and $A_{d_{min}}$, we show by construction that if we apply the precoding or pre-transformation on the same same length and rate profile, in a one to one mapping of min-weight polar codewords to the corresponding PAC codewords, some of the min-weight codewords may find a larger weight as a result of precoding. This mapping is shown in Fig. \ref{fig:mapping} where only a portion of min-weight codewords on the left hand side (i.e. polar codebook) are mapped to the collection of min-weight codewords in PAC codes on the right hand side, shown by arrow (i). 


Let us consider all the min-weight rows in $\mathbf{P}_n$ as a subset of all the min-weight codewords of polar codes. 
If we attempt to produce such min-weight codewords in PAC coding, we shall see that as a result of convolutional precoding $\mathbf{v}\mathbf{G}$, 1) some of these codewords are kept unchanged, 2) some are replaced with a different min-weight codewords, and 3) some are replaced with codewords with larger weights.

Note that in polar coding, $\mathbf{v}=\mathbf{u}$ as there is no precoding operation. Now, consider a row $\mathbf{g}_i$ of $\mathbf{P}_n$ with $w(\mathbf{g}_i)=w_{min}$ as a minimum weight codeword of the polar code.  In order to generate such a codeword, a vector $\mathbf{u}$  such that $u_i=1$ and $u_j=0$ for $j\neq i$, is needed to have $\mathbf{u}\mathbf{P}_n=\mathbf{g}_i$. However, such a vector $\mathbf{u}$ may not be obtained by precoding   $\mathbf{v}\mathbf{G}=\mathbf{u}$
since $\mathbf{v}$ contains frozen bits with coordinates in $\mathcal{A}^c$. 

Recall $u_j = \sum_{k=0}^m c_k v_{j-k}$ from Section \ref{sec:prelim}. In order to get $u_j=0$ for any $j>i$ and $j\in\mathcal{A}$, it is possible to choose either $v_j=0$ (for the case $\sum_{k=1}^m c_k v_{j-k}=0$) or 1 (when $\sum_{k=1}^m c_j v_{j-k}=1$). However, for any $j\in\mathcal{A}^c$, by convention $v_j=0$ in the rate profile. Hence, $u_j=1$ when $\sum_{k=1}^m c_k v_{j-k}=1$. This inevitably combines $\mathbf{g}_i$ with $\mathbf{g}_j$ for any $j\in\mathcal{A}^c$ where $u_j=1$. As Lemma \ref{lma:1} showed, the resulting weight will be

\begin{equation} \label{eq:comb1}
w(\mathbf{g}_i \oplus \bigoplus_{j\in \mathcal{J}\subseteq\mathcal{A}^c} \mathbf{g}_j) \geq w_{min} 
\end{equation} 
where $\mathcal{J}=\{j|j\in\mathcal{A}^c, j>i, \text{and } u_j=1\}$.

Now, we look at the three aforementioned resulting cases:
\begin{enumerate}
    \item $\mathbf{g}_i \oplus \bigoplus_{j\in \mathcal{J}\subseteq\mathcal{A}^c} \mathbf{g}_j=\mathbf{g}_i$: This case occurs where there is no $j\in\mathcal{A}^c$ for $j>i$ (i.e., $\mathcal{J}=\emptyset$) or depending on the choice of polynomial $\mathbf{g}$, we may get $u_j=0$ for any $j\in\mathcal{A}^c$ and $j>i$.
    \item $\mathbf{g}_i \oplus \bigoplus_{j\in \mathcal{J}\subseteq\mathcal{A}^c} \mathbf{g}_j=\mathbf{x}$ where $\mathbf{x}\neq\mathbf{g}_i$ but $w(\mathbf{x})=w_{min}$: This case occurs where $w(\bigoplus_{j\in \mathcal{J}\subseteq\mathcal{A}^c} \mathbf{g}_j)=2w(\mathbf{g}_i\wedge\bigoplus_{j\in \mathcal{J}\subseteq\mathcal{A}^c} \mathbf{g}_j)$ as according to the principle of inclusion exclusion, we have
    \begin{equation}\label{eq:wt2}
    \begin{split}
    w(\mathbf{g}_i &\oplus \bigoplus_{j\in \mathcal{J}\subseteq\mathcal{A}^c} \mathbf{g}_j)=w(\mathbf{g}_i)+\\&w(\bigoplus_{j\in \mathcal{J}\subseteq\mathcal{A}^c} \mathbf{g}_j)-2w(\mathbf{g}_i\wedge\bigoplus_{j\in \mathcal{J}\subseteq\mathcal{A}^c} \mathbf{g}_j)
\end{split}
\end{equation}
The operator wedge product $\wedge$ is equivalent to bit-wise ANDing.
    \item $\mathbf{g}_i \oplus \bigoplus_{j\in \mathcal{J}\subseteq\mathcal{A}^c} \mathbf{g}_j=\mathbf{x}$ where $\mathbf{x}\neq\mathbf{g}_i$ and $w(\mathbf{x})>w_{min}$: This case occurs where $w(\bigoplus_{j\in \mathcal{J}\subseteq\mathcal{A}^c} \mathbf{g}_j)>2w(\mathbf{g}_i\wedge\bigoplus_{j\in \mathcal{J}\subseteq\mathcal{A}^c} \mathbf{g}_j)$

\end{enumerate}
The second case is where the min-weight codewords in PAC codes differ from the ones in polar codes, however they are still min-weight codewords. The third case is where PAC codes lose some of the min-weight codewords that exist in polar codes. 
Note that the resulting larger weight codewords change the weight distribution of PAC codes.

In similar way, we can show that this event occurs for the minimum weight codewords resulting from the combination of more than one row of $\mathbf{P}_n$ with indices in set $\mathcal{A}$. 

\begin{example} \label{ex:1}
For the polar code and PAC code of (64,48,4) with RM-polar rate profile, we have $A_{4}=432$ and 320 for polar codes and PAC codes, respectively. The set $\mathcal{M}=\{i|i\in\mathcal{A},\text{and }  w(\mathbf{g}_i)=w_{min}\}=\{20,24,34,36,40,48\}$ and the set $\mathcal{N}=\{j|j\in\mathcal{A}^c,\text{and for any } i\in\mathcal{M},  j>i\}=\{32,33\}$.  Assuming $\mathbf{c}=[1, 0, 1, 1, 0, 1, 1]$, then instead of $\mathbf{g}_{20}$, we will have $\mathbf{g}_{20}\oplus\mathbf{g}_{32}$ yet with weight $w_{min}=4$ (case 2) in the codebook of PAC codes. Note that the elements of vector $\mathbf{u}$ are zeros except at coordinates 20 and 32, however, the vector $\mathbf{v}$ will have many non-zero elements in order to get the aforementioned vector $\mathbf{u}$ after precoding. 
Also,instead of $\mathbf{g}_{24}$, we will have $\mathbf{g}_{24}\oplus\mathbf{g}_{33}$ with weight 6 which is greater than $w_{min}$ (case 3 shown by arrow (ii) in Fig. \ref{fig:mapping}). The other rows with min-weight including $\mathbf{g}_{34}$, $\mathbf{g}_{36}$, $\mathbf{g}_{40}$, and $\mathbf{g}_{48}$ will exist unchanged in the codebook of PAC codes as there is no row $j\in\mathcal{A}^c$ for $j>34$ (case 1).
\end{example}

Note that by applying the precoding, there is no way to generate min-weight codewords other than based on the coset $C(0_0^{i-1},1)$ where $w(\mathbf{g}_i)=w_{min}$ as the following corollary concludes.
\begin{corollary} \label{cor:1}
If $i\in\mathcal{A}$ and $w(\mathbf{g}_i)>w_{min}$, inclusion of row(s) $k\in\mathcal{A}^c$ with $k>i$ in the coset  $C(0_0^{i-1},1)$ does not produce a coset with weight $w_{min}$ or less.
\end{corollary}
\begin{proof}
It follows directly from Lemma \ref{lma:1} the weight of the coset $C(0_0^{i-1},1)$ cannot be smaller than the weight of $\mathbf{g}_i$.
\end{proof}

Now, consider the codewords with weight larger than $w_{min}$ resulting from 
the coset $C(0_0^{i-1},1)$ where  $w(\mathbf{g}_i)=w_{min}$. The inclusion of $\mathbf{g}_j$ for $j>i$ and $j\in\mathcal{A}^c$  in the coset as a result of precoding may reduce the weight of some of the corresponding codewords in the polar codes. This case is shown by arrow (iii) in Fig. \ref{fig:mapping}.

\begin{example} \label{ex:3}
For the polar code and PAC code of (32,16,8) with RM rate profile, the set $\mathcal{M}=\{i|i\in\mathcal{A},\text{and }  w(\mathbf{g}_i)=w_{min}\}=\{13,14,21,22,25,26,28\}$ and the set $\mathcal{N}=\{j|j\in\mathcal{A}^c,\text{and for any } i\in\mathcal{M},  j>i\}=\{16,17,18,20,24\}$.  Considering the codeword resulting from the combination $\mathbf{g}_{13}\oplus\mathbf{g}_{22}$ which gives the weight $w(\mathbf{g}_{13}\oplus\mathbf{g}_{22})=12$, by inclusion of $\mathbf{g}_{18}$ ($j=18\in\mathcal{A}^c$), the weight will be $w(\mathbf{g}_{13}\oplus\mathbf{g}_{22}\oplus\mathbf{g}_{18})=8$.
\end{example}

One can observe that statistically the case of getting the weight $w_{min}$ or any specific weight as a result of inclusion of the rows in $\mathcal{A}^c$ is less frequent relative to the case of getting a weight larger than $w_{min}$. The numerical results of enumeration of min-weight codewords support this observation.  As Table \ref{tbl:1} shows the reduction in the min-weight codewords of polar codes except in an special case which is the subject of the following corollary.

\begin{corollary} \label{cor:2}
Suppose $\mathcal{M}=\{i|i\in\mathcal{A},\text{and }  w(\mathbf{g}_i)=w_{min}\}$. If for any $i\in\mathcal{M}$, there is no $j\in\mathcal{A}^c$ such that $j>i$,
then 
$A_{d_{min}}(\mathbf{v}\mathbf{G}\mathbf{P}_n) = A_{d_{min}}(\mathbf{v}\mathbf{P}_n)$.
\end{corollary}
\begin{proof}
In this case, there is no inclusion of rows with index $j\in\mathcal{A}^c$ in the coset $C(0_0^{i-1},1)$ where $w(\mathbf{g}_i$. Hence, as it was discussed earlier, it is possible to find a vector $\mathbf{v}$ to generate all the possible combinations of rows identical to polar codes.
\end{proof}

\begin{example} \label{ex:2}
For the polar code and PAC code of (128,32,16) with RM-polar rate profile, we have $A_{16}=56$. Knowing  $\mathcal{M}=\{114,116,120\}$, for any $i\in\mathcal{M}$, there is no $j\in\mathcal{A}^c$ such that $j>i$ as the largest $j$ in $\mathcal{A}^c$ is 113.
\end{example}

In summary, the inevitable inclusion of row(s) $\mathbf{g}_j$ for any $j\in \mathcal{A}^c$ and $j>i$ 
to the row combinations which are supposed to give min-weight codewords in polar codes may result in codewords with larger weights (arrow (ii) in Fig. \ref{fig:mapping}). Note that this inclusion depends on $\mathbf{g}$ and $\mathbf{d}$ and here we just discussed the possibility of the inclusion in general, regardless of the choice of $\mathbf{g}$ which is discussed in the next section.




The inclusion of frozen rows occurs in PC-Polar codes and polar codes with dynamic-frozen bits as well although there is no analytical explanation in the literature about the reason of their improvement.  In fact, the inclusion of frozen rows are a sabotage in the process of formation of minimum weight codewords.



\section{Weaknesses of Convolutional Precoding:\\Unequal Error Protection (UEP) and Lack of Cross-segmental Convolution}
We observed in Section \ref{sec:weight_reduction} that precoding in PAC codes can reduce $A_{d_{min}}$. It is difficult to systematically design a generator polynomial $\mathbf{g}$ that provides the minimum $A_{d_{min}}$. Nonetheless, we can design the precoding stage to mitigate the potential weakness or shortcoming of convolutional precoding. To do so, we  look at the precoding as a protection means for information bits similar to convolutional codes. We also consider the  This weakness is due to a relatively short constraint length.

\begin{figure}[t]
    \centering
    \includegraphics[width=1\columnwidth]{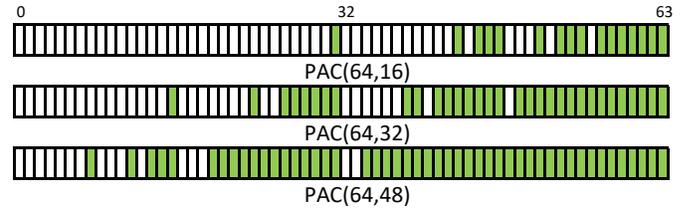}
    \caption{RM-polar rate-profiles for block-length $N=64$ and code rates $R=1/4,1/2,3/4$. 
    Green cells are members of set $\mathcal{A}$.} 
    \label{fig:rate_profiles}
\end{figure}

\begin{figure}
    \centering
    \includegraphics[width=0.7\columnwidth]{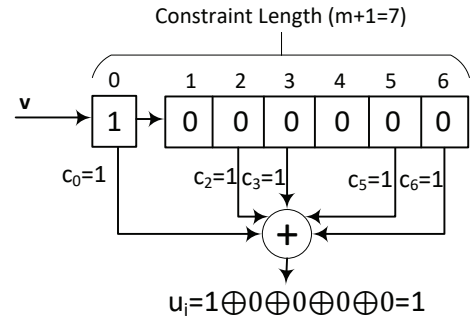}
    \caption{An example of  convolution in practice}
    \label{fig:conv2}
\end{figure}

Let us first study the distribution of the elements of set $\mathcal{A}$ in the rate-profile. Although the rate-profiles can be constructed with different methods \cite{harish}, here we consider the RM-Polar rate-profile \cite{li} which performs better on short codes. In this rate-profile, the weight of the rows in $\mathbf{P}_n$, denoted by $w(\mathbf{g}_j)$ for row $j$, plays an important role. As the code rate increases, $d_{min}=2^{w(j)}$, where $w(j)$ is the weight of binary expansion of $j$, increases. That leaves gaps between the bits in the set $\mathcal{A}$ and excludes the rows with weights lower than $d_{min}$. Fig. \ref{fig:rate_profiles} illustrates the gaps with white cells. When it comes to the convolution operation, these gaps makes the error protection of a subset of $\mathcal{A}$ weaker than the rest of the bits. Let us observe this weakness by an example. Consider bit $i=38$ in PAC(64,32). Since $v_i=0$ for any $i\in[32,37]$, if the constraint length $m$ is $m\leq 6$, then $u_{38} = \sum_{j=0}^m g_j v_{i-j}=v_{38}$. As you may notice, no convolution is happening here. In fact, the effective generator polynomial for $i=38$ and 39 is $\mathbf{g}=[0,...,0]$.  As a result, the bit $i=38$ which turns out to be transmitted over a relatively low-reliability sub-channel is left unprotected. Fig. \ref{fig:conv2} illustrates the case where $u_i=v_i$ as the shift-register is empty.  Note that we do not face this issue in the convolutional codes as there is no prefixed zero values in the input sequence to the encoder. 


This weakness may be mitigated by a longer constraint length and a proper generator polynomial $\mathbf{g}$ or by a different convolution scheme that has a longer memory. The longer constraint length requires a longer memory size $m$ for each path in the list decoding. A recommended long-memory polynomial is
\begin{equation}
  \mathbf{g}=[\mathbf{g}_{(i)}|0,\ldots,0|\mathbf{g}_{(ii)}]  
\end{equation}
where $\mathbf{g}_{(i)}$ and $\mathbf{g}_{(ii)}$ are the coefficients of two generator polynomials. The sub-sequence zeros in the concatenation helps in combination of {\em cross-segmental bits}. Segments are equilength ordered set of bits  which are obtained by dividing a block code $x_0^{N-1}$ into $M$ sub-blocks of size $2^m$ bits where $M=N/2^m=2^{n-m}$. The cross-segmental protection may help in the prevention of the correct path elimination, in particular when the error occurs in a segment in the middle and not in the first segment. 

\begin{figure}[t]
    \centering
    \includegraphics[width=0.7\columnwidth]{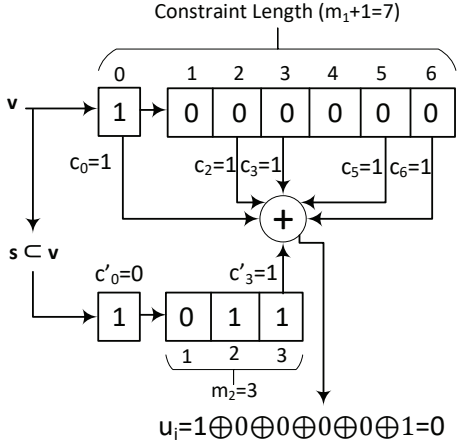}
    \caption{A different scheme to mitigate the effect of unequal error protection with two generator polynomial $\mathbf{g}_{(a)}=[1,0,1,1,0,1,1]$ and $\mathbf{g}_{(b)}=[0,0,0,1]$.} 
    \label{fig:new_cov}
\end{figure}

A smarter scheme that provides a longer memory without a large memory requirement is the scheme shown in Fig. \ref{fig:new_cov}. In this scheme, we add another shift register in parallel with the main shift-register, where we store a subset of input $\mathbf{v}$ stream, preferably the bits transmitted through low-reliability sub-channels. Note that the number of low-reliability bits in each segment is limited. Since the secondary shift-register has lower number of inputs, a subset of $\mathbf{v}$, the bits remains in the shift-register for a longer time-steps. This equivalent to having a longer memory.

The proposed schemes to improve the unequal error protection of the bits and to provide an additional cross-segmental protection will results in a fewer number of min-weight codewords comparing with conventional PAC codes. Table \ref{tbl:2} lists $A_{d_{min}}$ of some examples.

\begin{table}[h!]
\centering
\begin{tabular}{|c|c|} 
 \hline
  Polynomial & $\approx A_{16}$  \\ 
 \hline
  $\mathbf{g}=[1]$ & 94488  \\ 
 \hline
  $\mathbf{g}=[1,0,1,1]$ & 7520  \\ 
 \hline
  $\mathbf{g}=[1,0,1,1,{\color{blue}0,1,1}]$ & 3120   \\ 
 \hline
  $\mathbf{g}=[1,0,1,1{\color{blue},0,1,1}{\color{purple},0,1,1}]$ & 2812   \\ 
 \hline
 \hline
  $\mathbf{g}=[1,0,1,1{\color{blue},0,1,1},0\stackrel{\times 20}{\cdots\cdots},1,1{\color{blue},0,1,1}]$ & 2556  \\
 \hline
  $\mathbf{g}_{(a)}=[1,0,1,1{\color{blue},0,1,1}]$ \& $\mathbf{g}_{(b)}=[0,0,1,1,0,1,1]$ & 2574  \\  
  \hline
\end{tabular}
\caption{The number of min-weight codewords, $A_{d_{min}}$, with RM rate profile for PAC code (128,64,16) under various precoding schemes. The polynomial $\mathbf{g}=[1]$ is equivalent to no precoding, hence the output of encoder is a polar code.}

\label{tbl:2}
\end{table}

Lets us discuss the advantage of these example polynomials. Since $v_i=0$ for any $i\in[16,22]$ and this is the longest sub-sequence of zeros in the rate-profile, the constraint length $m+1$ should be $m\geq 8$. The polynomial $\mathbf{g}=[1,0,1,1,0,1,1,0,1,1]$ is an example that mitigates the unequal error protection resulting in a smaller $A_{d_{min}}$. A short polynomial such as  $\mathbf{g}=[1,0,1,1]$ results in a larger $A_{d_{min}}$ for the same reason. Intuitively, one can observe that this increase  is due to less inclusion of rows of $\mathbf{P}_n$ corresponding to $v_i=0$ in the row combinations as discussed earlier. 
The polynomial $\mathbf{g}=[1,0,1,1,0,1,1,0\stackrel{\times 20}{\cdots\cdots},1,1,0,1,1]$ helps more in cross-segmental protection. We can also use a longer polynomial for $\mathbf{g}_{(i)}$ to get to improve it from UEP point of view. Lastly, the two polynomials $\mathbf{g}_{(a)}=[1,0,1,1,0,1,1]$ and $\mathbf{g}_{(b)}=[0,0,1,0,1]$ where $\mathbf{g}_{(b)}$ is used for a low-reliability subset of indices in $\mathcal{A}$. This scheme contributes in mitigating  UEP and providing cross-segmental protection to some extent.

\begin{figure}
    \centering
    \includegraphics[width=0.9\columnwidth]{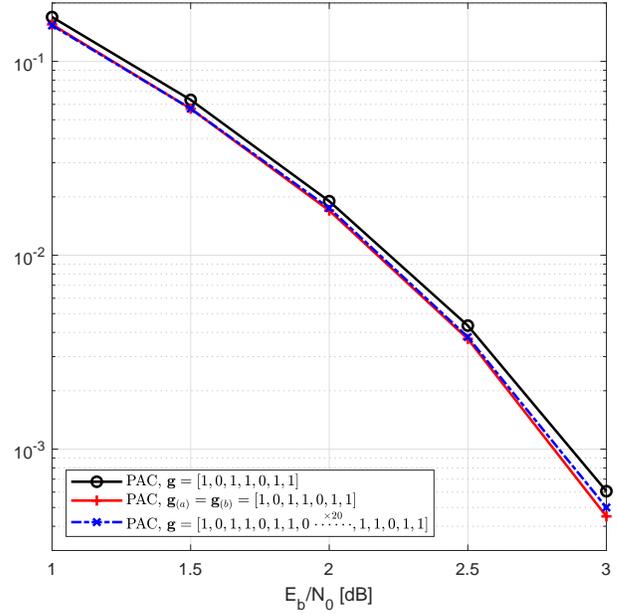}
    \caption{Performance of PAC code (128,64,16) with different precoding polynomials under list decoding with $L$=32.}
    \label{fig:FER}
\end{figure}

Fig. \ref{fig:FER} illustrates the FER performance of some of the polynomials listed in Table \ref{tbl:2}. As can be seen, the improvement is about 0.1 dB at high SNRs as we expect from the union bound $P_e^{ML} \approx A_{d_{min}} Q(\sqrt{2d_{min}R E_b/N_0})$ where  $A_{d_{min}}$ has reduced by the proposed irregular convolutions. 

\section{Conclusion} 
In this paper, we investigate the reason behind the reduction of the number of min-weight codewords in PAC codes. We also show where the precoding stage is  not effective depending on the code and the set $\mathcal{A}^c$. Then, we recognize the weaknesses of convolutional precoding and propose two approaches to mitigate them. 







\end{document}